\documentclass{article}[12pt]

\usepackage{geometry} % Adjust page margins
\usepackage{setspace} % Set line spacing
\usepackage{titlesec} % Format section titles
\usepackage[round]{natbib} % Bibliographic references (adjust options as per your citation style)
\usepackage{lipsum} % Generate dummy text (remove this package in your actual document)
\usepackage{pdfpages}
\usepackage{booktabs}
\usepackage{url}

\usepackage{breakurl}
\usepackage[breaklinks]{hyperref}
\titleformat{\section}{\normalfont\Large\bfseries}{\thesection}{1em}{}
\titleformat{\subsection}{\normalfont\large\bfseries}{\thesubsection}{1em}{}
\titleformat{\subsubsection}{\normalfont\normalsize\bfseries}{\thesubsubsection}{1em}{}

\title{Analyzing the Endeavours of the Supreme Court of India to Transcribe and Translate Court Arguments in Light of the Proposed EU AI Act }

\author{Kshitiz Verma\\ JK Lakshmipat University, Jaipur.\\ vermasharp@gmail.com}
\date{\today}

\begin{document}

\maketitle

\begin{abstract}
The Supreme Court of India has been a pioneer in using ICT in courts through its e-Courts project in India. Yet another leap, its recent project, \emph{Design, Development, and Implementation of Artificial Intelligence (AI) solution, tools for transcribing arguments and Court proceedings at Supreme Court of India}, expressed through bid number \href{https://main.sci.gov.in/pdf/TN/23052023_062513.pdf}{\bf Ref No. AI Solutions/2023/SCI} has potential to impact the way AI algorithms are designed in India, and not just for this particular project. In this paper, we evaluate the endeavours of the Supreme Court of India in light of the state of AI technology as well as the attempts to regulate AI. We argue that since the project aims to transcribe and translate the proceedings of the constitutional benches of the Supreme Court, it has potential to impact \emph{rule of law} in the country. Hence, we place this application in High Risk AI as per the provisions to the proposed EU AI Act. We provide some guidelines on the approach to transcribe and translate making the maximum use of AI in the Supreme Court of India without running into the dangers it may pose. 

\end{abstract}

\section{Introduction}

Indian judiciary is burdened by a heavy backlog of cases and it is in a desperate searchfor innovative methods to address the backlog. The use of technology has helped Indian judiciary to battle the situation \citep{verma2018courts}. The technology is growing at a very rapid pace and hence it should leverage new technologies like artificial intelligence. In a written reply in Rajya Sabha on 07 April 2022, the then Union Minister of Law and Justice, Shri Kiren Rijiju had highlighted various use cases of AI in judiciary in India \citep{kirenreply2022}. The Supreme Court of India has decided to do a pilot project on the use of AI for transcribing constitutional court arguments in real time \citep{scibid}. When this paper was written, the Supreme Court had already asked for participation of technology companies in a bid to use AI in the court proceedings.

\subsection{Our Focus on the Supreme Court Bid Document}

As the Supreme Court mentions in its bid document that the goal is to leverage artificial intelligence, machine learning and deep learning for various purposes including the automated transcription of the proceedings of the the Supreme Court's constitutional benches \citet{scibid}. The tasks that are of interest to us, and are discussed in the paper, are enumerated and for the sake of completeness, reproduced below. %We have also mentioned the sections in which these topics are dealt in detail in our paper.
\begin{enumerate}
\item Page 40, Section V(b)(i.) \emph{Transcribing of arguments and court proceedings on real time basis and displaying the same on monitors in the courtroom.} %(Section \ref{sec:transcribe})
\item Page 40, Section V(b)(iii.) \emph{The transcription generated primarily must be in English language. However, the transcription generated must also be capable of being translated into the languages stated in the Eight Schedule of the Constitution of India, 1950.} %(Section \ref{sec:translate})
\item Page 40, Section V(b)(iv.) \emph{The AI tool to be developed and deployed must have an advance level of natural language processing, to understand legal terms, documents, petitions, judgments, etc. and to automatically classify them in the relevant specialization.} 
\item Page 40, Section V(b)(v.) \emph{The AI tool to be developed and deployed must have software and machine learning capabilities, to build a sophisticated hierarchy of classification models to analyse the contents of documents transcribed contained in unstructured text, rich text, html, PDF documents, to have a prediction, intelligent processing, smart classification, content extraction and summarization.}

% \item Page 41, Section V(d) \emph{The SCI initially intends to deploy the AI tool in the court rooms where the Constitution Bench matters shall be listed for hearing.} %(Section \ref{sec:transcribe}, \ref{sec:translate})
% \item Page 41, Section V(f) \emph{The AI solution and tools will be hosted on the server(s) of the Supreme Court of India, which can be on-prem and/or cloud, and if the solution is server-less, then the same shall remain within the control of the Supreme Court of India. AI tool shall not use any external APIs, tools or packages developed by third parties, which are not free and open source, to avoid any dependency on its functioning.}
% \item Page 41, Section V(h) \emph{Audio-video recording of 05 cases will be provided to the Bidder, to enable understanding the assignment.}  %(Section \ref{sec:transcribe}, \ref{sec:translate})
% \item Page 42, Section V(m) \emph{The AI tools and solution shall be operationalized on turnkey basis i.e., within 60 days from the date of issuance of purchase order (PO)/letter of intent (LoI).}  %(Section \ref{sec:transcribe})

\end{enumerate}

AI is a constantly evolving scientific area right now. There is no consensus on any single definition of AI or even threats that it may pose \citep{hinton_threat}, \citep{yoshua_threat} \citep{lecun_threat}, \citep{andrew_threat}. In such times, it may be said that it is very courageous of the Supreme Court to invite bids for using AI to improve access to justice in India. However, these endeavours have to be properly evaluated on legal and technical basis so as to appropriately balance the need of the judiciary to use AI and the robustness of the technology to be fit for the purpose.

% \begin{enumerate}
\subsection{Elements of High Risk AI in the Supreme Court Project}

The early and landmark step of the European Union to frame a law for the use of artificial intelligence is commendable that has been lauded throughout the world. The proposed EU AI  Act of the \citet{EUAIAct} considers that the AI systems that adversely impact the following may be classified as high risk AI:
% \subsection{Yardstick of risk}
\begin{enumerate}
\item Health and safety of natural persons
\item Adverse impact on fundamental rights of natural persons
\item Democracy and rule of law
\item Environment
\end{enumerate}

The intention of the Supreme Court is to use AI for constitutional bench matters, that are binding on courts all over India, including the Supreme Court benches that have a strength lesser than the said constitutional bench. Since the jurisdiction of Supreme Court constitutional benches is really supreme, it may directly impact health and safety of the citizens if a case with it is about health or any other rights, including fundamental rights. It may impact the environment if the case is of such nature. Even if all such impacts may look far fetched, being used in constitutional benches, use of AI may certainly impact rule of law. In any case, playing the devil's advocate and assuming a worst case scenario will only help the Supreme Court to assess the gravity of application of AI in its proceedings. Such an application of AI, therefore, may be placed under the high risk category as defined in the proposed EU AI Act.

Thus, the stakes are high if the Supreme Court wants to utilize the capabilities of AI for transcribing and translating constitutional benches proceedings as any mistake made by the AI may really lead to a blunder. For this reason, borrowing some procedures for conformity assessment from the proposed EU AI Act may be helpful.

\subsection{Anil Ambani's Case}

Anil Ambani's case \citep{anilambanicase} is an excellent example of sensitivity of words used in the Supreme Court judgments and how AI may catastrophically change meanings. Transcribing of voice may not be a big deal when a machine learning algorithm on YouTube tries to generate the transcripts of a random video. However, when it is utilized by the Supreme Court of India, it no longer remains an ordinary application. It has potential to touch millions, if not billions of lives in India. We all are aware of the infamous case in which the word ``NOT" was removed from the order and removal of this one word had resulted in two staff members being terminated from their services \citep{anilambanicase}.

\subsubsection {Reasonably Foreseeable Misuse}

A wrong transcription may have a tremendous impact on rule of law. Human beings may suffer from automation bias. This may mean that the burden of correctness may be placed on the technology. Thus, even if the error should have been caught by human beings, it may creep in due to automation bias. However, since the transcriptions will be displayed on screens in the court, someone should be able to figure out such mistakes.

\subsubsection {Malicious Manipulation}
As above, a wrong transcription may still be acceptable as it may be corrected in the better versions or by a careful use of the technology. However, if the technology is weaker in cybersecurity and it is possible to manipulate the AI models maliciously, then it might be easy to attribute a deliberate malicious manipulation to wrong transcription by the model. This may lead to more cases like above and may make the error more difficult to attribute to a human being.

\section{Use of AI in Judiciary in Other Jurisdictions}

%
% \begin{enumerate}
%
% \item Legislation on AI of various jurisdictions like the EU, Australia, USA and UK.
% \item Comparative studies done on implementation of AI in high risk areas.
% \item Has any other judicial wing of any other democratic country tried to use it?
% \item What do the scholars and the scientists have to say about the use of AI in law, judiciary and law enforcement?
%
%
% \end{enumerate}

Many researchers in USA have found use cases of applications of AI in judicial system where too much discretionary power is provided to the courts. For example in granting bails and asylums. Humans are known to suffer from various biases and such biases often reflect while granting bail and asylums. In such cases use of algorithms may be better \citep{sunstein2021governing} \citep{jung2020simple}\citep{kleinberg2018human}. Many of these proposals are being considered for improving access to justice in such scenarios. After a yearlong trial, Charlotte's jail population is down almost 20 percent without increasing crime. At the same time, use of AI in policing has been criticized by many researchers \citep{castelvecchi2020mathematicians}. In the UK, Durham HART model that was developed jointly by Durham Constabulary and the University of Cambridge, has been criticized because of adverse impacts on society and legal perspectives \citep{oswald2018algorithmic}.

Singapore courts have been announcing to use AI in automatic transcription of court proceedings and a use of case of legal area classification is discussed in \citet{howe2019legal}.

We do not focus on providing an exhaustive list here, a more comprehensive discussion on such related work may be found in \citet{vidhiAIpolicy}.

\section{State of AI for various Applications}

AI is not mature yet. It is still evolving and a steady state has not been achieved so as to say something about its use definitively. One recent example is degradation in the performance of GPT-4 on some tasks  \citep{chen2023chatgpts}. AI's applications in judiciary may suffer from bias, discrimination, lack of transparency, loss of autonomy of judicial actors \citep{leslie2021artificial}.

The intentions of the Supreme Court may be right but the technology is not yet there where it can be seamlessly used in the Supreme Court for transcribing or translation. More so when the environment in the court may be noisy or the goal may be to translate documents to all the languages recognized by the Constitution of India. Hence, any project carried out in a period of 60 days is certainly going to fall short of the quality that the hon'ble court is looking for \citep{scibid}. The  solution and solution provider must undergo certain conformity assessments to ensure that the system works the way it is alleged to be working and all foreseeable risks have been recognized, acknowledged, mitigated and eliminated. The model should have some degree of explainability than being a black box like most of the AI models currently are. Hence, the Supreme Court should create a framework for the companies solving such complex problems. We are afraid that the current bid document of the Supreme Court of India is not specific enough to provide guidelines on the design of the AI to the solution providers.

The Supreme Court can be a torch bearer in the development of explainable AI in India. If Supreme Court puts forth a requirement of explainability, it might be an inspiration for many other government agencies to do so, allowing integration of AI in mainstream governance in a swift manner mitigating the risks that AI may pose.

We now discuss the state of AI in the two applications that are mentioned in the Supreme Court bid.

\subsection{AI in Relation to Transcription of Proceedings}

The speech recognition technology is relatively late to take off compared to vision and natural language processing. One of the revolutionary ideas in automatic speech recognitions have been implemented by \citet{baevski2020wav2vec}. This advance has enabled scaling up to 1,000,000 hours of training data \citep{zhang2022bigssl}. Even in YouTube, that has got a lot of videos to train on, accuracy is lesser than desired \citep{lin2022developing}. One of the state of the art models of speech recognition, developed by OpenAI -- Whisper, was trained on 680,000 hours of data \citep{radford2023robust}. It still has an average word error rate (WER) of 12\% on various speech datasets. In a more curated Large Vocabulary Continuous Speech Recognition (LVCSR) dataset the errror rates are still around 6\%. Automatic speech recognition (ASR), like other AI, is also known to exhibit bias \citep{martin2022bias}\citep{feng2021quantifying}. Thus, we have enough mainstream research literature from automatic speech recognition to show that the technology is far from being 100\% accurate. This calls for evaluation of the transcription done by any system, particularly, if it is for judicial use.

Different accents add further complexities to the problem of automatic speech recognition \citep{viglino2019end}, \citep{prasad2020accents}, \citep{hinsvark2021accented}. The judges in the Supreme Court may come from different states and may have different accents. Moreover, the accent of Indians is different compared to those in English speaking nations. This comparison is important because the most of the success of Automatic Speak Recognition is from western English datasets and such datasets are baised against the non-native speakers. Hence, the Supreme Court proceeding is a very particular scenario, much different from usual transcription scenarios. The datasets required should be from such distribution only. Hence, it is imperative for the Supreme Court to provide not just 5 videos but as many as are available so that a good training dataset may be prepared which may be further used by researchers and companies alike. AI models are important but so is data \citep{sambasivan2021everyone}.

\subsection{AI in Relation to Translating Documents}
The success of natural language processing and machine translation has seen amazing results for English, French, Spanish, Japanese, etc. However, it is not the same for all languages in the world. \citet{joshi-etal-2020-state} study the disparity between languages from the point of view of research conducted and resources created for processing. This has created a big technological divide between the languages. They categorize languages in six classes, class 5 being the maximally resource rich and class 0 being the least. Table \ref{tab:list} provides the list of the languages provided in the Eighth Schedule of the Constitution of India along with the class they belong to as per \citet{joshi-etal-2020-state}. As per classification of languages based on the datasets and labels available, the success of class 5 languages is less likely to be replicated for other classes . There is no Indian language in class 5. Hindi and other Indian languages fall in class 4 or lower. This means that datasets as well as labels for these languages are several orders of magnitude smaller compared to English. Thus, the same success stories are not likely to be sung for the translation of English documents to the languages recognized by the Constitution of India. 

Hence, one has to be really careful translating legal texts. Translating legal documents comes with their own pros and cons. While translations enable a wider access of the legal provisions, use of various words and their interpretations in other languages may create confusions too. However, the pros clearly outweigh cons, it is worth doing it but again some guidelines and checks have to be put in.

%Any institutions, including judiciary, using it in the EU or anywhere else?

%Translation is a well studied problem. Is there enough data for legal use there?

%Talk of how even the human generated texts have created problems in the UN system. How texts in English and French have been interpreted differently?

\begin{table}[h]
    \centering
%     \rotatebox{270}{%
    \begin{tabular}{|c|c|}
        \hline
        {\bf Language} & {\bf Class} \\
        \hline
        Assamese & 1 \\
        \hline
        Bengali & 3 \\
        \hline
        Gujarati & 1 \\
        \hline
        Hindi & 4 \\
        \hline
        Kannada & 1 \\
        \hline
        Kashmiri & 1 \\
        \hline
        Konkani & 2 \\
        \hline
        Malayalam & 1 \\
        \hline
        Manipuri & 1 \\
        \hline
        Marathi & 2 \\
        \hline
        Nepali & 1 \\
        \hline
        Oriya & 1 \\
        \hline
        Punjabi & 2 \\
        \hline
        Sanskrit & 2 \\
        \hline
        Sindhi & 1 \\
        \hline
        Tamil & 3 \\
        \hline
        Telugu & 1 \\
        \hline
        Urdu & 3 \\
        \hline
        Bodo & 0 \\
        \hline
        Santhali & 1 \\
        \hline
        Maithili & 1 \\
        \hline
        Dogri & 0 \\
        \hline
    \end{tabular}
%     }
    \caption{Class of 22 languages recognized by the Eighth Schedule of the Constitution of India as per \citet{joshi-etal-2020-state}. Higher class means higher availability of data that may enable a high quality processing.}
    \label{tab:list}
\end{table}

\subsection{Other Complex Legal Tasks}
 
The bid document mentions to build systems that understand legal terms, documents, petitions, judgments, etc. and to automatically classify them in the relevant specialization. The bid document also aims to build a sophisticated hierarchy of classification models to analyze the contents of documents transcribed contained in unstructured text, rich text, html, PDF documents, to have a prediction, intelligent processing, smart classification, content extraction and summarization.

%\subsubsection {Have complex legal understanding} to 

AI systems have no ``understanding" as such. It has even sparked a philosophical debate on what understanding even means. While classification is certainly a doable task, preparing training datasets for it is time consuming and the exact problems that need to be solved also needs to be defined first. The current text in the bid document is not clear. So, we do not address these issues in the current paper and focus on transcription and translation only.

\section{Guidelines for the Implementation Inspired by the Proposed EU AI Act}

The stakes are high as constitutional benches are going to use AI in their proceedings for transcription and translation. Any mistake, if goes unnoticed, may lead to a blunder. Hence, we need to minimize errors in the output of the AI system. For this, we need to follow some rules. Our further analysis is inspired by the proposed EU AI Act \citep{EUAIAct}.  Thus, the application of AI in the Supreme Court Constitutional Bench proceedings may be classified as a high risk AI that may impact rule of law. For this reason, some procedures for conformity of high risk AI may be borrowed from the proposed EU AI Act. The core principles for AI systems as provided in the proposed EU AI Act are reproduced below:
\begin{enumerate}
\item Human agency and oversight
\item Technical robustness and safety
\item Privacy and data governance
\item Transparency
\item Diversity, non-discrimination and fairness
\item Social and environmental well-being
\end{enumerate}

A typical machine learning application has an average cycle of around 6-12 months. Hence, our first suggestion is that instead of making it as a two months project for a solution provider, it should be much longer project for the Supreme Court itself. In this paper, by adhering to the principles enshrined in the proposed EU AI Act that are specific for the requirements of the Supreme Court, our goal is to provide some guiding principles to the solution provider that is going to implement the AI project.  We enumerate the required conformity assessment below as provided from Articles 9 to 15 of the proposed EU AI Act. 

\subsection{Risk Management by the Solution Provider} A risk management system is a continuous iterative process that runs throughout the entire life cycle of a high-risk AI system. In our context, it means until the system for automatic transcription of the Supreme Court arguments is in place, it should be monitored for its output. Once the application becomes so robust that it looses high risk status, vigilance may be reduced. The idea is to monitor for identification, estimation and evaluation of the known and the reasonably foreseeable risks that the high risk AI system can pose. There is clearly a reasonably foreseeable risk associated with the applications of AI for transcribing and translation. We have also seen it before that just omitting one word may lead to catastrophes and Automatic Speech Recognition systems are known to error on words while creating transcriptions, i.e., their transcriptions are far from being 100\% accurate. Hence, our suggestions are as follows. 
\begin{enumerate}
\item There must be a continuous monitoring and evaluation of the AI system during development by the solution provider, as well as for the whole duration it is in use by the Supreme Court.  
\item The specific risks that may be posed are identified, estimated and evaluated. In this case, the system may transcribe words wrongfully, insert or delete words. Estimation of such errors needs to be done well and the solution must be evaluated for improving. This may create situations like the Anil Ambani case. In case of translations, wrong translations may create legal discrepancies for interpretation which may create more problems than solving the existing ones.
\item Look for effective ways to eliminate, mitigate or reduce the risks identified. Improve the model and the training data. One of the best ways is to put the output for the public inspection, which the Supreme Court is already doing by running the transcriptions on the screens in the court. For translations, it may upload unreliable copies and ask the relevant legal fraternity to check for the quality of translation and help improve the quality of the training dataset by providing better translations.  
%\item Test for intended purpose and compliance.
\end{enumerate}

\subsection{Data and Data Governance Followed by the Solution Provider} There must be a check on the quality of data being used for training the transcription and translation AI systems. The Supreme Court must ensure that the data used for training, validation and testing are appropriate for the intended purpose of the system. The data used should not exceed what is required for training the desired system and should not devoid the solution provider of necessary data. There must be a reasonable attempt to remove all kinds of biases from the system. The biases may be due to region, accent, gender or any other basis. 

\begin{enumerate}
\item The foremost thing for the solution provider is to prepare a dataset that is worthy of solving the problem reasonably well. For the modern deep learning systems, the bigger the dataset, the better. However, the quality also plays a crucial role. Transcribing and translation are both supervised learning tasks, so we need the quality and labelled datasets for both. Given that these tasks may pose high risk to the rule of law, the solution provider needs to pay special attention to the quality of the datasets prepared and it should have the same environment as a typical Supreme Court Constitutional Bench proceeding has. 

\item The data should be clearly defined for the purpose at hand, which in this case is transcribing and translation.
\item There is a live streaming of various high courts too. Even that data may be used for preparing training dataset for transcriptions. 
\item For translations, the high courts may also be asked to prepare translations of the judgments in the respective regional language so that the issues of less data for parallel corpus of English with Indian languages may be addressed. 
%\item The applications need to be careful as to minimize the bias in the system.
\end{enumerate}

\subsection{Transparency of Machine Learning Operations} The said transcription and translation AI systems should be designed and developed in such a way to ensure that their operations are sufficiently transparent to enable the solution provider and the Supreme Court and its registry to reasonably understand the system's functioning in accordance with the intended purpose of the AI system.

% Transparency shall mean that all technical means available in accordance with the generally acknowledged latest state of the art used to ensure that the AI system's output is interpretable by the provider and the user. The level of accuracy, robustness and cybersecurity against which the high risk AI system has been tested and validated and which can be expected, and any clearly known and foreseeable circumstances that may have an impact on that expected level of accuracy, robustness and cybersecurity. In case of known foreseeable circumstances of misuse which may lead to impact on fundamental rights or rule of law.
The degree to which the AI system can provide an explanation for decisions it takes should also be maximized. In the present case, the explanations for why some sound was transcribed as a particular word is important. Also, why a particular translation was chosen for a particular sentence is important. Thus, relevant information about the Supreme Court actions that may influence performance, including the type or quality of input data should be included. Any necessary maintenance and care measures to ensure the proper functioning of that AI system should also be deployed.

\begin{enumerate}
\item The Supreme Court is already planning to show the transcriptions so generated on the screens in the courtroom on the real-time basis. This is a really good step for enforcing the solution provider towards transparent machine learning operations, as the AI system's performance may be seen by the legal fraternity and any errors may be readily found.
\item It may be a even better idea to use colours to interpret the output of the system. For example, if the system has a high confidence in generating some transcription, it may colour it in black, output with intermediate confidence may be coloured yellow and outputs with low confidence may be coloured red. It may highlight deletions by underscores. It will help the readers to catch errors effortlessly. However, automation bias may creep in as the text with high confidence of AI may get less attention from the human readers than it deserves. A similar colouring scheme may be used for translation too. 
\item Evaluations should be transparent. The AI system needs to understand if its performance changes as a function of changes in judge, accent, gender, region, etc. This should be taken as input in designing the AI systems. 
\end{enumerate}

\subsection{Record keeping by the AI System} The AI system should keep a record of the way it is working by automatic recording of events (`logs'). This is to ensure the traceability of the way AI system is functioning.
% Those logging capabilities shall conform to the state of the art and recognised standards or common specifications.
\begin{enumerate}
\item The application should be designed to provide extensive logs for useful and important events. For example, it is good to keep logs for such instances in which the systems output is marked as underscore, yellow or red. This enables to explain better the way the algorithm is behaving.
\item In the translation process, each sentence must be assigned a low, intermediate or high confidence and assigned colours as for the transcription process. The exact way in which the confidence may be assigned will depend on the implementation of the AI model. 
\end{enumerate}

\subsection{Human Oversight} As the risk posed by an AI system increases, increased amount of human oversight is required, proportionate to the risks associated with those systems. Natural persons in charge of ensuring human oversight need to have sufficient level of AI literacy and the necessary support and authority to exercise that function and to allow for thorough investigation, if required. Human oversight shall aim to minimize the risks to health, safety, fundamental rights and rule of law. Human oversight shall take into account the specific risks, the level of automation, and context of the AI system. This is to avoid automation bias so that we do not consider the output of the system as correct without questioning or until a damage occurs.
\begin{enumerate}
\item The Supreme Court should appoint persons with sufficient AI literacy for this purpose. Their task will be just to evaluate the functioning of the AI sytems. 
\item The solution provider should explicitly assign the task of human oversight to its team implementing the solution.
\end{enumerate}

\subsection{Cyber Security} Since it has impact on the rule of law in India, it is very sensitive from cybersecurity perspective. Such systems are sensitive to attacks as changing just one word may lead to catastrophes and cyber criminals may try to change the transcriptions deliberately by attacking the AI model and its parameters. 
\begin{enumerate}
\item Security should be built in design by default and the security standards should be equivalent to financial institutions like banks may be put in force. 
\item There should be a due diligence on making systems secure by keeping a dedicated team of cyber-security experts from either industry or academia.
\item The system should be resilient to an unauthorized change of the model parameters or any faults or inconsistencies.
\item Once a system is compromised, it may be easy to disguise a deliberate act as a mistake made by the AI model. Hence, the solution should take steps to minimize malicious manipulation.
\end{enumerate}

\subsection{Technical Documentation Accompanying the Solution}
\label{subsec:td}

A technical documentation showing that all the provisions regarding the risk management, data governance, transparency of machine learning operations,  record keeping, human oversight and cybersecurity are met should be a mandatory deliverable by the solution provider before putting the AI system in use by the Supreme Court. It should provide information on all the principles mentioned above. It should include a general description of the AI system including:

\begin{enumerate}
%\item its intended purpose,
\item The nature of the data likely or intended to be processed by the system.
\item The description of the hardware on which the AI system is intended to run.
\item A detailed and easily intelligible description of the system's main optimization goal or goals.
\item A detailed and easily intelligible description of the system's expected output and expected output quality.
\item A detailed and easily intelligible instructions for interpreting the system's output. Report errors in the output of the AI in black/yellow/red transcriptions mentioned before.
\item Some examples of scenarios for which the system should not be used. An explicit declaration from the solution provider if no such scenario exists. 
\item A detailed description of the elements of the AI system and the process for its development, including:
\begin{enumerate}
\item methods and steps performed for the development of the AI system,
\item a description of the architecture, design specifications, algorithms and the data structures, how they related to one another and provide the overall processing of the AI system,
\item a description of the data obtained, labelling procedures, data cleaning methodologies, etc.,
\item assessment of human oversight, i.e., a detailed report on how human oversight helped in mitigating the errors that otherwise may creep in, 
\item validation and testing procedures used, including information about the validation and testing data used and their main characteristics, metrics used to measure accuracy, robustness, etc., 
\item cybersecurity measures put in place, does deploying this AI exposes the SCI or its website to cybercriminals in any manner? 
\end{enumerate}
%\item A detailed information about the monitoring, functioning and control of the AI system, in particular with regard to its capabilities and limitations in performance , including the degree of accuracy for specific persons or group of persons on which the system is intended to be used and the overall expected level of accuracy in relation to 
\item A description of the appropriateness of the performance metrics for the specific AI system.
\item A detailed description of the risk management system.
\item A report should be made public for the inspection of academics and civil society.
\end{enumerate}

\section{Conclusion}
This paper presents the state of the art principles to be applied on AI systems whose application is in sensitive areas like judiciary. We conclude that the pilot project taken up by the Supreme Court of India to implement AI for transcribing the Supreme Court constitutional bench proceedings is commendable but a bit earlier in time. The success of the project largely depends on the datasets prepared for the said task. We also discussed the use of AI to translate legal texts from English to other Indian languages. We concluded that it is even more difficult task to do such translation.  In our opinion, the Supreme Court of India may take up a long term project for creation of datasets which may be made available to the public at large. This will foster research. This will certainly take more than 60 days, as required by the Supreme Court in the current bid document but it will be a much more robust solution. Finally, we conclude by suggesting that the company providing the solution must be asked to provide details and a technical report as mentioned in Section \ref{subsec:td} should be submitted. 

% References
% \section*{References}
\bibliographystyle{plainnat} % Choose your desired citation style
\bibliography{references} % Replace with the name of your .bib file containing the references

\appendix

% Scope of Work included
% \includepdf[pages={40-45}]{../bid.pdf}
\end{document}